\documentclass[aip,jcp,reprint]{revtex4-1}
\usepackage[T1]{fontenc}
\usepackage[latin9]{inputenc}
\usepackage{units}
\usepackage{amstext}
\usepackage{graphicx}
\usepackage{esint}
\usepackage{lipsum}

\begin{document}

\title{Molecular diffusion between walls with adsorption and desorption}

\author{Maximilien Levesque}
\email{maximilien.levesque@gmail.com}
\affiliation{CNRS, UPMC Univ. Paris 06, UMR 7195 PECSA, 75005 Paris, France}

\author{Olivier B\'enichou}
\affiliation{CNRS, UPMC Univ. Paris 06, UMR 7600 LPTMC, 75005 Paris, France}

\author{Benjamin Rotenberg}
\affiliation{CNRS, UPMC Univ. Paris 06, UMR 7195 PECSA, 75005 Paris, France}

\begin{abstract}
The time dependency of the diffusion coefficient of particles in porous media is an efficient probe of their geometry. The analysis of this quantity, measured \emph{e.g.} by nuclear magnetic resonance (PGSE-NMR), can provide rich information pertaining to porosity, pore size distribution, permeability and surface-to-volume ratio of porous materials. Nevertheless, in numerous if not all practical situations, transport is confined by walls where adsorption and desorption processes may occur. In this article, we derive explicitly the expression of the time-dependent diffusion coefficient between two confining walls in the presence of adsorption and desorption. We show that they strongly modify the time-dependency of the diffusion coefficient, even in this simple geometry. We finally propose several applications, from sorption rates measurements to the use as a reference for numerical implementations for more complex geometries.
\end{abstract}

\maketitle

\section{Introduction}

The time dependency of the diffusion coefficient of particles in porous media is an efficient probe of their geometry\cite{mitra_diffusion_1997,sen_time-dependent_2003,sen_geometry_2004,dudko_berez_2005,valfouskaya_adler_2005}. Extracting information on the microscopic structure of the porous medium from diffusion measurements, has seen applications from heterogeneous geological materials, \emph{e.g.} clays\cite{valfouskaya_adler_2005} to biological tissues, \emph{e.g.} brain\cite{brain_1992}. In this respect, pulsed gradient spin echo nuclear magnetic resonance measurements (PGSE-NMR), unraveled by explicit derivations of the time-dependent diffusion coefficient of tracers in simple, model media,\cite{dudko_berez_2005,berezhkovskii_biased_2010,berezhkovskii_analytical_2011} have shown to reveal rich information regarding to porosity, pore size distribution, permeability and surface-to-volume ratio.\cite{kleinberg_reviewNMRporous_1999}

Nevertheless, in numerous if not all practical situations, transport is confined by walls where 
adsorption and desorption processes may occur. This is especially true for porous media for which the surface-to-volume ratio is large.
Such processes dramatically alter the overall dynamics of the diffusing species, even in the most simple geometries\cite{alcor_PNAS_2004,levitz_intermittent_2008,benichou_optimal_2010,metzler_effective_2011,levesque_taylor_2012,dudko_berez_2005}. 

Here, we derive explicitly the expression of the time-dependent diffusion coefficient between two confining walls in the presence of general adsorption and desorption.

\section{Results and Discussion}

Following a stochastic approach, we consider a Brownian particle moving along the $x$ axis with a bulk
diffusion coefficient $D_{b}$, confined between two walls at
positions $x=0$ and $x=L$. The evolution equations and the boundary
conditions are 
\begin{eqnarray}
\partial_{t}c\left(x,t\right) & = & D_{b}\frac{\partial^{2}c}{\partial x^{2}}\text{ for \ensuremath{x\in\left]0,L\right[}},\label{eq:def1}\\
\dot{\Gamma}_{0}(t) & = & -k_{d}\Gamma_{0}+k_{a}c\left(0,t\right)=D_{b}\left.\partial_{x}c\right|_{x=0},\label{eq:def2}\\
\dot{\Gamma}_{L}(t) & = & -k_{d}\Gamma_{L}+k_{a}c\left(L,t\right)=-D_{b}\left.\partial_{x}c\right|_{x=L},\label{eq:def3}
\end{eqnarray}
where $c\left(x,t\right)$ {[}length$^{-3}${]} is the particle concentration
at position $x$ and time $t$, $\Gamma_{0}$ and $\Gamma_{L}$ {[}length$^{-2}${]}
are the surface concentrations in adsorbed particles at both walls. The desorption and adsorption rates are noted 
$k_d$ {[}time$^{-1}${]} and $k_a$ {[}length$\cdot$time$^{-1}${]}. The brownian particles can be in three states. In state $a_{b}$, they're free to move along the $x$ axis. In states $a_{0}$ or $a_{L}$, they're adsorbed at one wall at $x=0$ or $L$, where they remain immobile.
The equilibrium probability~$P_{\text{eq}}$ associated to these states follows from the normalization condition:
\begin{equation}
\sum_{\alpha=a_{0},a_{L}}P_{\text{eq}}\left(\alpha\right)+\int_{0}^{L}P_{\text{eq}}\left(a_{b}\right)\text{d}x=1,
\end{equation}
with the result $P_{\text{eq}}\left(a_{0}\right)=P_{\text{eq}}\left(a_{L}\right)=k_{a}/(k_dL+2k_a)$
and $P_{\text{eq}}\left(a_{b}\right)=k_d/(k_dL+2k_a)$.

Assuming that the particles are initially distributed according to $P_\text{eq}$, the mean squared displacement (MSD), $M(t)=\left\langle \left(x\left(t\right)-x_{0}\right)^{2}\right\rangle$, of
the particles can be expressed as
\begin{widetext}
\begin{eqnarray}
M(t) = \int_{0}^{L}\text{d}x\int_{0}^{L}\text{d}x_{0}\left(x-x_{0}\right)^{2}\sum_{\alpha,\beta}P_{\text{eq}}\left(\alpha\right)G\left(x,\beta,t|x_{0},\alpha\right),\label{eq:msd}
\end{eqnarray}
\end{widetext}
where $G\left(x,\beta,t|x_{0},\alpha\right)$ is the particle propagator,
\emph{i.e.} the probability for a particle initially at $x_0$ in state $\alpha$ to be at $x$ in state $\beta$ at time $t$.
Several propagators have to be calculated, which correspond to all combinations of starting and ending positions and states.
Their explicit determination is conveniently performed by first Laplace transforming Eq.~\ref{eq:def1}--\ref{eq:def3}. In the Laplace domain, boundary conditions associated with adsorption and desorption simplify to radiative boundary conditions (see, for example, Ref.~\cite{berezhkovskii_escape_2009,levesque_taylor_2012}). Then, after standard calculations, the Laplace transform of these propagators are found to be given by:

\begin{widetext}
\begin{eqnarray}
\widetilde{G}\left(x<x_{0},a_{b},s|x_{0},a_{b}\right) & = & \kappa q(\gamma\cosh(qx)+q\sinh(qx))\left(\gamma\cosh\left(q\left(x_{0}-L\right)\right)-q\sinh\left(q\left(x_{0}-L\right)\right)\right),\\
\widetilde{G}\left(x>x_{0},a_{b},s|x_{0},a_{b}\right) & = & \kappa q\left(\gamma\cosh\left(qx_{0}\right)+q\sinh\left(qx_{0}\right)\right)(\gamma\cosh(q(x-L))-q\sinh(q(x-L))),\\
\widetilde{G}\left(x,a_{0},s|x_{0},a_{b}\right) & = & \frac{\kappa}{k_{a}}q\left(\gamma k_{a}-s\right)(\gamma\cosh(q(L-x))+q\sinh(q(L-x))),\\
\widetilde{G}\left(x,a_{L},s|x_{0},a_{b}\right) & = & \frac{\kappa}{k_{a}}q\left(\gamma k_{a}-s\right)(\gamma\cosh(qx)+q\sinh(qx)),\\
\widetilde{G}\left(x,a_{b},s|x_{0},a_{0}\right) & = & \frac{\kappa}{k_{d}}q\left(\gamma k_{a}-q^{2}D_{b}\right)\left(\gamma\cosh\left(q\left(x_{0}-L\right)\right)-q\sinh\left(q\left(x_{0}-L\right)\right)\right),\\
\widetilde{G}\left(x,a_{0},s|x_{0},a_{0}\right) & = & \frac{\kappa}{k_{a}k_{d}}q\left(\gamma k_{a}-s\right)\left(\left(\gamma k_{a}+q^{2}D_{b}\right)\cosh(Lq)+q\left(k_{a}+\gamma D_{b}\right)\sinh(Lq)\right),\\
\widetilde{G}\left(x,a_{L},s|x_{0},a_{0}\right) & = & \frac{\kappa}{k_{a}k_{d}}\left(\gamma k_{a}-s\right)\left(\gamma qk_{a}-q^{3}D_{b}\right),\\
\widetilde{G}\left(x,a_{b},s|x_{0},a_{L}\right) & = & \frac{\kappa}{k_{d}}q\left(\gamma k_{a}-q^{2}D_{b}\right)\left(\gamma\cosh\left(qx_{0}\right)+q\sinh\left(qx_{0}\right)\right),\\
\widetilde{G}\left(x,a_{0},s|x_{0},a_{L}\right) & = & G\left(x,a_{L},t|x_{0},a_{0}\right),\\
\widetilde{G}\left(x,a_{L},s|x_{0},a_{L}\right) & = & G\left(x,a_{0},t|x_{0},a_{0}\right),
\end{eqnarray}
\end{widetext}

where $s$ is the conjugate of $t$, $q=\sqrt{s/D_b}$, $\gamma=\displaystyle \frac{k_d + s}{k_a}$ and $\kappa=\left(s\left(\gamma^{2}+q^{2}\right)\sinh(Lq)+2s\gamma q\cosh(Lq)\right)^{-1}$.

\begin{figure}
\includegraphics[width=8.5 cm]{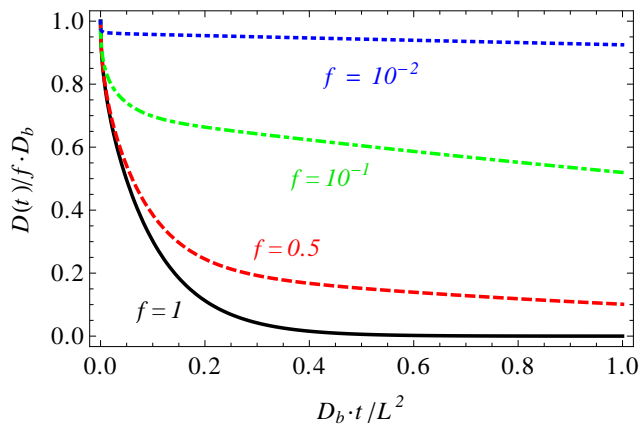}

\caption{Time-dependent diffusion coefficient in reduced units as a function
of the reduced time. Several ratios $f=(1+2k_a/k_dL)^{-1}$ are reported, which decrease with the strength of the sorption
phenomena: (black line) $f=1$, \emph{i.e.} without adsorption; (red dashed line)
$f=0.5$; (green dot-dashed line) $f=10^{-1}$ and (blue dotted line) $f=10^{-2}$.\label{fig:diffusion_wrt_t}}
\end{figure}

Inserting these propagators and their respective $P_{\text{eq}}$ in Eq.~\ref{eq:msd}, one gets for the Laplace transform of the mean squared displacement:

\begin{widetext}
\begin{equation}
\widetilde{M}(s) = \frac{1}{2k_{a}+k_{d}L}
\times\left[\frac{2k_{d}L}{sq^{2}}\right.\left.\frac{4k_{d}(k_{d}+s)\text{sinh}\left(\frac{q L}{2}\right)}{sq^{3}\left((k_{d}+s)\text{cosh}\left(\frac{q L}{2}\right)+k_{a}q~\text{sinh}\left(\frac{q L}{2}\right)\right)}\right]\label{eq:msds},
\end{equation}
\end{widetext}

where $s$ is the conjugate variable of time $t$ in Laplace space, and $q=\sqrt{\frac{s}{D_{b}}}$. The diffusion coefficient
is then easily derived from the MSD as 
\begin{equation}
D(t) = \mathcal{L}^{-1}\left( \frac{s}{2}\widetilde{M}(s) \right),\label{eq:diffusion}
\end{equation}
where $\mathcal{L}^{-1}$ is the inverse Laplace transform.
The resulting reduced time-dependent
diffusion coefficient is plotted in figure~\ref{fig:diffusion_wrt_t}
as a function of the reduced time $D_{b}t/L^{2}$, for various adsorption and
desorption rates\cite{abate_laplace_2004}.
At $t=0$, $D(0)/D_{b}$ is simply the fraction $f=(1+2k_a/k_dL)^{-1}$ of mobile (non-adsorbed) particles.
Figure~\ref{fig:diffusion_wrt_t} explores a large range of $f$,
from weak to strong sorption. For all sorption parameters
$k_{a}$ and $k_{d}$, diffusivity decreases with time, as increases
the probability for each Brownian particle that explores the media to experience confinement
by a wall and immobilization by adsorption. This decrease is strongly modified by the sorption processes, as the more probable it is for a particle to get adsorbed, the more time it takes to explore the media.
At long times, because our media is not permeable, \emph{i.e.} because the confinement is total in our geometry, the effective diffusion coefficient tends toward zero.

\section{Short and Long times Approximations}

\begin{figure}
\includegraphics[width=8.5 cm]{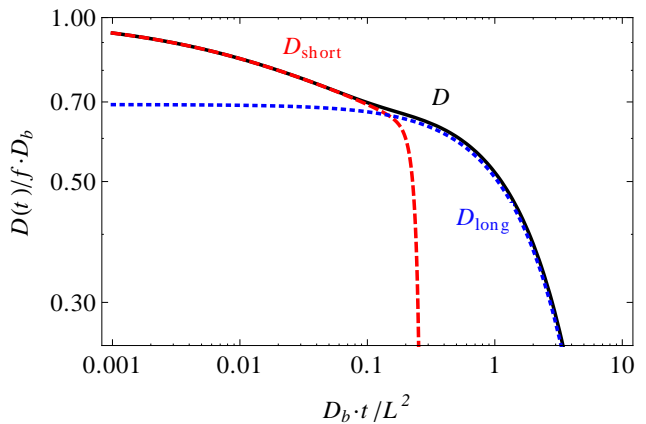}

\caption{Reduced diffusion coefficient as a function of reduced time for $f=10^{-1}$;
(black) from the full expression of Eq.~\ref{eq:diffusion}, (red
dashed line) from the short time approximation of Eq.~\ref{eq:diff_short_times} up to order 10
and (blue dotted line) from the long time approximation of Eq.~\ref{eq:diff_long_times}.\label{fig:limits}}
\end{figure}

For $\frac{D_{b}}{k_{d}L^{2}}+\frac{k_{a}}{2k_{d}L}\gg1$, \emph{i.e.} for sufficiently long residence time on walls and fraction of adsorbed
species, the series expansion of the hyperbolic functions in Eq.~\ref{eq:msds} allows, together with 
 the residue theorem, to explicitely inverse Eq.~\ref{eq:diffusion} in the long time-limit (\emph{i.e.} for $s\rightarrow0$), leading to:
\begin{equation}
\frac{D_{\text{long}}\left(t\right)}{D_{b}}=f \times \frac{k_{a}L}{2D_{b}+k_{a}L}e^{\displaystyle-\frac{k_{d}}{1+\frac{k_{a}L}{2D_{b}}}t}.\label{eq:diff_long_times}
\end{equation}
At short times, we analytically inverse the power series expansion
for $\mathcal{L}^{-1}\left[\lim_{s\rightarrow\infty}D\left(s\right)\right]$.
The first terms read:
\begin{equation}
\frac{D_{\text{short}}(t)}{D_b} = f\times\left(1-\frac{4}{\sqrt{\pi}}\frac{\sqrt{D_bt}}{L} +2\frac{k_at}{L}\right) +\mathcal{O}\left(t^\frac{3}{2}\right).
\label{eq:diff_short_times}
\end{equation}
See Supplementary Material Document No.~\cite{si} for terms up to order 10 of the expansion.
Figure~\ref{fig:limits} compares the short and long time approximations of
the time-dependent diffusion coefficient to the full numerical inversion of the Laplace transform
of Eq.~\ref{eq:diffusion} for $f=10^{-1}$. Together, the short and long time approximations
provide a simple uniform approximation over the whole temporal domain.

\section{Conclusion}

In summary, we addressed explicitely the effect of adsorption
and desorption phenomena on the time-dependent diffusion coefficient
of Brownian particles confined between walls. The full expression
of $D(t)$ is given, together with convenient and accurate approximations. It is shown that adsorption and desorption processes strongly modify $D(t)$, so that their consideration is important for proper determination of micro-geometric information from $D(t)$ measurements.
Eq.~\ref{eq:diffusion} may also be used to extract sorption rates $k_a$ and $k_d$ from experimental measurements of $D(t)$, \emph{e.g.} by PGSE-NMR.
Finally, in more complex geometries, resort to numerical simulations such as Lattice Boltzmann\cite{rotenberg_europhysics_2008,rotenberg_faraday_2010} is necessary.
The present work provides exact reference results required for the validation of such numerical schemes in the presence of adsorption and desorption.

\begin{acknowledgments}
The authors thank D. Frenkel and I. Pagonabarraga for useful discussions. BR and ML acknowledge financial support from the Agence Nationale de la Recherche under grant ANR-09-SYSC-012 and OB acknowledges support from  European Research Council starting Grant FPTOpt-277998.
\end{acknowledgments}

\bibliographystyle{apsrev4-1}

\begin{thebibliography}{19}%
\makeatletter
\providecommand \@ifxundefined [1]{%
 \@ifx{#1\undefined}
}%
\providecommand \@ifnum [1]{%
 \ifnum #1\expandafter \@firstoftwo
 \else \expandafter \@secondoftwo
 \fi
}%
\providecommand \@ifx [1]{%
 \ifx #1\expandafter \@firstoftwo
 \else \expandafter \@secondoftwo
 \fi
}%
\providecommand \natexlab [1]{#1}%
\providecommand \enquote  [1]{``#1''}%
\providecommand \bibnamefont  [1]{#1}%
\providecommand \bibfnamefont [1]{#1}%
\providecommand \citenamefont [1]{#1}%
\providecommand \href@noop [0]{\@secondoftwo}%
\providecommand \href [0]{\begingroup \@sanitize@url \@href}%
\providecommand \@href[1]{\@@startlink{#1}\@@href}%
\providecommand \@@href[1]{\endgroup#1\@@endlink}%
\providecommand \@sanitize@url [0]{\catcode `\\12\catcode `\$12\catcode
  `\&12\catcode `\#12\catcode `\^12\catcode `\_12\catcode `\%12\relax}%
\providecommand \@@startlink[1]{}%
\providecommand \@@endlink[0]{}%
\providecommand \url  [0]{\begingroup\@sanitize@url \@url }%
\providecommand \@url [1]{\endgroup\@href {#1}{\urlprefix }}%
\providecommand \urlprefix  [0]{URL }%
\providecommand \Eprint [0]{\href }%
\providecommand \doibase [0]{http://dx.doi.org/}%
\providecommand \selectlanguage [0]{\@gobble}%
\providecommand \bibinfo  [0]{\@secondoftwo}%
\providecommand \bibfield  [0]{\@secondoftwo}%
\providecommand \translation [1]{[#1]}%
\providecommand \BibitemOpen [0]{}%
\providecommand \bibitemStop [0]{}%
\providecommand \bibitemNoStop [0]{.\EOS\space}%
\providecommand \EOS [0]{\spacefactor3000\relax}%
\providecommand \BibitemShut  [1]{\csname bibitem#1\endcsname}%
\let\auto@bib@innerbib\@empty
\bibitem [{\citenamefont {Mitra}(1997)}]{mitra_diffusion_1997}%
  \BibitemOpen
  \bibfield  {author} {\bibinfo {author} {\bibfnamefont {P.~P.}\ \bibnamefont
  {Mitra}},\ }\href@noop {} {\bibfield  {journal} {\bibinfo  {journal} {Physica
  A: Stat. Mech. App.}\ }\textbf {\bibinfo {volume} {241}},\ \bibinfo {pages}
  {122} (\bibinfo {year} {1997})}\BibitemShut {NoStop}%
\bibitem [{\citenamefont {Sen}(2003)}]{sen_time-dependent_2003}%
  \BibitemOpen
  \bibfield  {author} {\bibinfo {author} {\bibfnamefont {P.~N.}\ \bibnamefont
  {Sen}},\ }\href {\doibase doi:10.1063/1.1611477} {\bibfield  {journal}
  {\bibinfo  {journal} {J. Chem. Phys.}\ }\textbf {\bibinfo {volume} {119}},\
  \bibinfo {pages} {9871} (\bibinfo {year} {2003})}\BibitemShut {NoStop}%
\bibitem [{\citenamefont {Sen}(2004)}]{sen_geometry_2004}%
  \BibitemOpen
  \bibfield  {author} {\bibinfo {author} {\bibfnamefont {P.~N.}\ \bibnamefont
  {Sen}},\ }\href {\doibase 10.1002/cmr.a.20017} {\bibfield  {journal}
  {\bibinfo  {journal} {Concepts in Magnetic Resonance}\ }\textbf {\bibinfo
  {volume} {{23A}}},\ \bibinfo {pages} {1} (\bibinfo {year}
  {2004})}\BibitemShut {NoStop}%
\bibitem [{\citenamefont {Dudko}\ \emph {et~al.}(2005)\citenamefont {Dudko},
  \citenamefont {Berezhkovskii},\ and\ \citenamefont
  {Weiss}}]{dudko_berez_2005}%
  \BibitemOpen
  \bibfield  {author} {\bibinfo {author} {\bibfnamefont {O.~K.}\ \bibnamefont
  {Dudko}}, \bibinfo {author} {\bibfnamefont {A.~M.}\ \bibnamefont
  {Berezhkovskii}}, \ and\ \bibinfo {author} {\bibfnamefont {G.~H.}\
  \bibnamefont {Weiss}},\ }\href {\doibase 10.1021/jp051172r} {\bibfield
  {journal} {\bibinfo  {journal} {J. Phys. Chem. B}\ }\textbf {\bibinfo
  {volume} {109}},\ \bibinfo {pages} {21296} (\bibinfo {year}
  {2005})}\BibitemShut {NoStop}%
\bibitem [{\citenamefont {Valfouskaya}\ \emph {et~al.}(2005)\citenamefont
  {Valfouskaya}, \citenamefont {Adler}, \citenamefont {Thovert},\ and\
  \citenamefont {Fleury}}]{valfouskaya_adler_2005}%
  \BibitemOpen
  \bibfield  {author} {\bibinfo {author} {\bibfnamefont {A.}~\bibnamefont
  {Valfouskaya}}, \bibinfo {author} {\bibfnamefont {P.~M.}\ \bibnamefont
  {Adler}}, \bibinfo {author} {\bibfnamefont {J.-F.}\ \bibnamefont {Thovert}},
  \ and\ \bibinfo {author} {\bibfnamefont {M.}~\bibnamefont {Fleury}},\ }\href
  {\doibase 10.1063/1.1871352} {\bibfield  {journal} {\bibinfo  {journal} {J.
  App. Phys.}\ }\textbf {\bibinfo {volume} {97}},\ \bibinfo {pages} {083510}
  (\bibinfo {year} {2005})}\BibitemShut {NoStop}%
\bibitem [{\citenamefont {LeBihan}\ \emph {et~al.}(1992)\citenamefont
  {LeBihan}, \citenamefont {Turner}, \citenamefont {Douek},\ and\ \citenamefont
  {Patronas}}]{brain_1992}%
  \BibitemOpen
  \bibfield  {author} {\bibinfo {author} {\bibfnamefont {D.}~\bibnamefont
  {LeBihan}}, \bibinfo {author} {\bibfnamefont {R.}~\bibnamefont {Turner}},
  \bibinfo {author} {\bibfnamefont {P.}~\bibnamefont {Douek}}, \ and\ \bibinfo
  {author} {\bibfnamefont {N.}~\bibnamefont {Patronas}},\ }\href@noop {}
  {\bibfield  {journal} {\bibinfo  {journal} {Amer. J. Roentgenol.}\ }\textbf
  {\bibinfo {volume} {159}},\ \bibinfo {pages} {591} (\bibinfo {year}
  {1992})}\BibitemShut {NoStop}%
\bibitem [{\citenamefont {Berezhkovskii}\ and\ \citenamefont
  {Dagdug}(2010)}]{berezhkovskii_biased_2010}%
  \BibitemOpen
  \bibfield  {author} {\bibinfo {author} {\bibfnamefont {A.~M.}\ \bibnamefont
  {Berezhkovskii}}\ and\ \bibinfo {author} {\bibfnamefont {L.}~\bibnamefont
  {Dagdug}},\ }\href {\doibase doi:10.1063/1.3489375} {\bibfield  {journal}
  {\bibinfo  {journal} {J. Chem. Phys.}\ }\textbf {\bibinfo {volume} {133}},\
  \bibinfo {pages} {134102} (\bibinfo {year} {2010})}\BibitemShut {NoStop}%
\bibitem [{\citenamefont {Berezhkovskii}\ and\ \citenamefont
  {Dagdug}(2011)}]{berezhkovskii_analytical_2011}%
  \BibitemOpen
  \bibfield  {author} {\bibinfo {author} {\bibfnamefont {A.~M.}\ \bibnamefont
  {Berezhkovskii}}\ and\ \bibinfo {author} {\bibfnamefont {L.}~\bibnamefont
  {Dagdug}},\ }\href {\doibase doi:10.1063/1.3567187} {\bibfield  {journal}
  {\bibinfo  {journal} {J. Chem. Phys.}\ }\textbf {\bibinfo {volume} {134}},\
  \bibinfo {pages} {124109} (\bibinfo {year} {2011})}\BibitemShut {NoStop}%
\bibitem [{\citenamefont {Kleinberg}(1996)}]{kleinberg_reviewNMRporous_1999}%
  \BibitemOpen
  \bibfield  {author} {\bibinfo {author} {\bibfnamefont {R.~L.}\ \bibnamefont
  {Kleinberg}},\ }\href@noop {} {\emph {\bibinfo {title} {Encyclopedia of
  Nuclear Magnetic Resonance vol.9}}},\ edited by\ \bibinfo {editor}
  {\bibnamefont {{Wiley, New-York}}}\ (\bibinfo {year} {1996})\ pp.\ \bibinfo
  {pages} {4960--4969}\BibitemShut {NoStop}%
\bibitem [{\citenamefont {Alcor}\ \emph {et~al.}(2004)\citenamefont {Alcor},
  \citenamefont {Croquette}, \citenamefont {Jullien},\ and\ \citenamefont
  {Lemarchand}}]{alcor_PNAS_2004}%
  \BibitemOpen
  \bibfield  {author} {\bibinfo {author} {\bibfnamefont {D.}~\bibnamefont
  {Alcor}}, \bibinfo {author} {\bibfnamefont {V.}~\bibnamefont {Croquette}},
  \bibinfo {author} {\bibfnamefont {L.}~\bibnamefont {Jullien}}, \ and\
  \bibinfo {author} {\bibfnamefont {A.}~\bibnamefont {Lemarchand}},\ }\href
  {http://www.jstor.org/stable/3372177} {\bibfield  {journal} {\bibinfo
  {journal} {PNAS}\ }\textbf {\bibinfo {volume} {101}},\ \bibinfo {pages}
  {8276} (\bibinfo {year} {2004})}\BibitemShut {NoStop}%
\bibitem [{\citenamefont {Levitz}\ \emph {et~al.}(2008)\citenamefont {Levitz},
  \citenamefont {Zinsmeister}, \citenamefont {Davidson}, \citenamefont
  {Constantin},\ and\ \citenamefont {Poncelet}}]{levitz_intermittent_2008}%
  \BibitemOpen
  \bibfield  {author} {\bibinfo {author} {\bibfnamefont {P.}~\bibnamefont
  {Levitz}}, \bibinfo {author} {\bibfnamefont {M.}~\bibnamefont {Zinsmeister}},
  \bibinfo {author} {\bibfnamefont {P.}~\bibnamefont {Davidson}}, \bibinfo
  {author} {\bibfnamefont {D.}~\bibnamefont {Constantin}}, \ and\ \bibinfo
  {author} {\bibfnamefont {O.}~\bibnamefont {Poncelet}},\ }\href {\doibase
  10.1103/PhysRevE.78.030102} {\bibfield  {journal} {\bibinfo  {journal} {Phys.
  Rev. E}\ }\textbf {\bibinfo {volume} {78}},\ \bibinfo {pages} {030102}
  (\bibinfo {year} {2008})}\BibitemShut {NoStop}%
\bibitem [{\citenamefont {B\'enichou}\ \emph {et~al.}(2010)\citenamefont
  {B\'enichou}, \citenamefont {Grebenkov}, \citenamefont {Levitz},
  \citenamefont {Loverdo},\ and\ \citenamefont
  {Voituriez}}]{benichou_optimal_2010}%
  \BibitemOpen
  \bibfield  {author} {\bibinfo {author} {\bibfnamefont {O.}~\bibnamefont
  {B\'enichou}}, \bibinfo {author} {\bibfnamefont {D.}~\bibnamefont
  {Grebenkov}}, \bibinfo {author} {\bibfnamefont {P.}~\bibnamefont {Levitz}},
  \bibinfo {author} {\bibfnamefont {C.}~\bibnamefont {Loverdo}}, \ and\
  \bibinfo {author} {\bibfnamefont {R.}~\bibnamefont {Voituriez}},\ }\href
  {\doibase 10.1103/PhysRevLett.105.150606} {\bibfield  {journal} {\bibinfo
  {journal} {Phys. Rev. Lett.}\ }\textbf {\bibinfo {volume} {105}},\ \bibinfo
  {pages} {150606} (\bibinfo {year} {2010})}\BibitemShut {NoStop}%
\bibitem [{\citenamefont {Chechkin}\ \emph {et~al.}(2011)\citenamefont
  {Chechkin}, \citenamefont {Zaid}, \citenamefont {Lomholt}, \citenamefont
  {Sokolov},\ and\ \citenamefont {Metzler}}]{metzler_effective_2011}%
  \BibitemOpen
  \bibfield  {author} {\bibinfo {author} {\bibfnamefont {A.~V.}\ \bibnamefont
  {Chechkin}}, \bibinfo {author} {\bibfnamefont {I.~M.}\ \bibnamefont {Zaid}},
  \bibinfo {author} {\bibfnamefont {M.~A.}\ \bibnamefont {Lomholt}}, \bibinfo
  {author} {\bibfnamefont {I.~M.}\ \bibnamefont {Sokolov}}, \ and\ \bibinfo
  {author} {\bibfnamefont {R.}~\bibnamefont {Metzler}},\ }\href {\doibase
  doi:10.1063/1.3593198} {\bibfield  {journal} {\bibinfo  {journal} {J. Chem.
  Phys.}\ }\textbf {\bibinfo {volume} {134}},\ \bibinfo {pages} {204116}
  (\bibinfo {year} {2011})}\BibitemShut {NoStop}%
\bibitem [{\citenamefont {Levesque}\ \emph {et~al.}(2012)\citenamefont
  {Levesque}, \citenamefont {B\'enichou}, \citenamefont {Voituriez},\ and\
  \citenamefont {Rotenberg}}]{levesque_taylor_2012}%
  \BibitemOpen
  \bibfield  {author} {\bibinfo {author} {\bibfnamefont {M.}~\bibnamefont
  {Levesque}}, \bibinfo {author} {\bibfnamefont {O.}~\bibnamefont
  {B\'enichou}}, \bibinfo {author} {\bibfnamefont {R.}~\bibnamefont
  {Voituriez}}, \ and\ \bibinfo {author} {\bibfnamefont {B.}~\bibnamefont
  {Rotenberg}},\ }\href {\doibase 10.1103/PhysRevE.86.036316} {\bibfield
  {journal} {\bibinfo  {journal} {Phys. Rev. E}\ }\textbf {\bibinfo {volume}
  {86}},\ \bibinfo {pages} {036316} (\bibinfo {year} {2012})}\BibitemShut
  {NoStop}%
\bibitem [{\citenamefont {Berezhkovskii}\ \emph {et~al.}(2009)\citenamefont
  {Berezhkovskii}, \citenamefont {Barzykin},\ and\ \citenamefont
  {Zitserman}}]{berezhkovskii_escape_2009}%
  \BibitemOpen
  \bibfield  {author} {\bibinfo {author} {\bibfnamefont {A.~M.}\ \bibnamefont
  {Berezhkovskii}}, \bibinfo {author} {\bibfnamefont {A.~V.}\ \bibnamefont
  {Barzykin}}, \ and\ \bibinfo {author} {\bibfnamefont {V.~Y.}\ \bibnamefont
  {Zitserman}},\ }\href {\doibase doi:10.1063/1.3160546} {\bibfield  {journal}
  {\bibinfo  {journal} {J. Chem. Phys.}\ ,\ \bibinfo {pages} {245104}}
  (\bibinfo {year} {2009})}\BibitemShut {NoStop}%
\bibitem [{\citenamefont {Abate}\ and\ \citenamefont
  {Valk\'o}(2004)}]{abate_laplace_2004}%
  \BibitemOpen
  \bibfield  {author} {\bibinfo {author} {\bibfnamefont {J.}~\bibnamefont
  {Abate}}\ and\ \bibinfo {author} {\bibfnamefont {P.~P.}\ \bibnamefont
  {Valk\'o}},\ }\href {\doibase 10.1002/nme.995} {\bibfield  {journal}
  {\bibinfo  {journal} {Int. J. Numer. Meth. Engng.}\ }\textbf {\bibinfo {volume} {60}},\ \bibinfo {pages} {979}
  (\bibinfo {year} {2004})}\BibitemShut {NoStop}%
\bibitem [{si()}]{si}%
  \BibitemOpen
  \href@noop {} {}\bibinfo {howpublished} {See supplementary material 1
  for terms up to order 10 of the short time expansion
  of the diffusion coefficient.}\BibitemShut {Stop}%
\bibitem [{\citenamefont {Rotenberg}\ \emph {et~al.}(2008)\citenamefont
  {Rotenberg}, \citenamefont {Pagonabarraga},\ and\ \citenamefont
  {Frenkel}}]{rotenberg_europhysics_2008}%
  \BibitemOpen
  \bibfield  {author} {\bibinfo {author} {\bibfnamefont {B.}~\bibnamefont
  {Rotenberg}}, \bibinfo {author} {\bibfnamefont {I.}~\bibnamefont
  {Pagonabarraga}}, \ and\ \bibinfo {author} {\bibfnamefont {D.}~\bibnamefont
  {Frenkel}},\ }\href {\doibase 10.1209/0295-5075/83/34004} {\bibfield
  {journal} {\bibinfo  {journal} {Europhysics Letters}\ }\textbf {\bibinfo
  {volume} {83}},\ \bibinfo {pages} {34004} (\bibinfo {year}
  {2008})}\BibitemShut {NoStop}%
\bibitem [{\citenamefont {Rotenberg}\ \emph {et~al.}(2010)\citenamefont
  {Rotenberg}, \citenamefont {Pagonabarraga},\ and\ \citenamefont
  {Frenkel}}]{rotenberg_faraday_2010}%
  \BibitemOpen
  \bibfield  {author} {\bibinfo {author} {\bibfnamefont {B.}~\bibnamefont
  {Rotenberg}}, \bibinfo {author} {\bibfnamefont {I.}~\bibnamefont
  {Pagonabarraga}}, \ and\ \bibinfo {author} {\bibfnamefont {D.}~\bibnamefont
  {Frenkel}},\ }\href {\doibase 10.1039/b901553a} {\bibfield  {journal}
  {\bibinfo  {journal} {Faraday Discussions}\ }\textbf {\bibinfo {volume}
  {144}},\ \bibinfo {pages} {223} (\bibinfo {year} {2010})}\BibitemShut
  {NoStop}%
\end{thebibliography}

%

\end{document}


\title{Supplementary Material for:\\ Molecular diffusion between walls with adsorption and desorption}

\author{Maximilien Levesque}
\email{maximilien.levesque@gmail.com}
\affiliation{CNRS, UPMC Univ. Paris 06, UMR 7195 PECSA, 75005 Paris, France}

\author{Olivier B\'enichou}
\affiliation{CNRS, UPMC Univ. Paris 06, UMR 7600 LPTMC, 75005 Paris, France}

\author{Benjamin Rotenberg}
\affiliation{CNRS, UPMC Univ. Paris 06, UMR 7195 PECSA, 75005 Paris, France}

\maketitle



\section{Short-time expansion of the time-dependent diffusion coefficient}

At short times $t$, the power series expansion for the time-dependent diffusion coefficient up to order~10 reads:

\scriptsize

\begin{eqnarray*}
\frac{D_\text{short}(t)}{f\cdot D_b} & = & 1 -\frac{4\sqrt{D_{b}}\sqrt{t}}{L\sqrt{\pi}} +\frac{2k_{a}t}{L}\\
 & &  -\frac{8k_{a}^{2}t^{3/2}}{3L\sqrt{\pi}\sqrt{D_{b}}} \\
 & & -\frac{k_{a}\left(D_{b}k_{d}-k_{a}^{2}\right)t^{2}}{LD_{b}}\\
 &  & +\frac{16k_{a}^{2}\left(2D_{b}k_{d}-k_{a}^{2}\right)t^{5/2}}{15L\sqrt{\pi}D_{b}^{3/2}}\\
 &  & +\frac{k_{a}\left(k_{a}^{4}-3D_{b}k_{d}k_{a}^{2}+D_{b}^{2}k_{d}^{2}\right)t^{3}}{3LD_{b}^{2}}\\
 &  & -\frac{32k_{a}^{2}\left(D_{b}k_{d}-k_{a}^{2}\right)\left(3D_{b}k_{d}-k_{a}^{2}\right)t^{7/2}}{105L\sqrt{\pi}D_{b}^{5/2}}\\
 &  & -\frac{k_{a}\left(-k_{a}^{6}+5D_{b}k_{d}k_{a}^{4}-6D_{b}^{2}k_{d}^{2}k_{a}^{2}+D_{b}^{3}k_{d}^{3}\right)t^{4}}{12LD_{b}^{3}}\\
 &  & +\frac{64k_{a}^{2}\left(2D_{b}k_{d}-k_{a}^{2}\right)\left(k_{a}^{4}-4D_{b}k_{d}k_{a}^{2}+2D_{b}^{2}k_{d}^{2}\right)t^{9/2}}{945L\sqrt{\pi}D_{b}^{7/2}}\\
 &  & +\frac{k_{a}\left(D_{b}k_{d}-k_{a}^{2}\right)\left(-k_{a}^{6}+6D_{b}k_{d}k_{a}^{4}-9D_{b}^{2}k_{d}^{2}k_{a}^{2}+D_{b}^{3}k_{d}^{3}\right)t^{5}}{60LD_{b}^{4}}\\
 &  & -\frac{128k_{a}^{2}\left(k_{a}^{4}-3D_{b}k_{d}k_{a}^{2}+D_{b}^{2}k_{d}^{2}\right)\left(k_{a}^{4}-5D_{b}k_{d}k_{a}^{2}+5D_{b}^{2}k_{d}^{2}\right)t^{11/2}}{10395L\sqrt{\pi}D_{b}^{9/2}}\\
 &  & -\frac{k_{a}\left(-k_{a}^{10}+9D_{b}k_{d}k_{a}^{8}-28D_{b}^{2}k_{d}^{2}k_{a}^{6}+35D_{b}^{3}k_{d}^{3}k_{a}^{4}-15D_{b}^{4}k_{d}^{4}k_{a}^{2}+D_{b}^{5}k_{d}^{5}\right)t^{6}}{360LD_{b}^{5}}\\
 &  & +\frac{256k_{a}^{2}\left(D_{b}k_{d}-k_{a}^{2}\right)\left(2D_{b}k_{d}-k_{a}^{2}\right)\left(3D_{b}k_{d}-k_{a}^{2}\right)\left(k_{a}^{4}-4D_{b}k_{d}k_{a}^{2}+D_{b}^{2}k_{d}^{2}\right)t^{13/2}}{135135L\sqrt{\pi}D_{b}^{11/2}}\\
 &  & +\frac{k_{a}\left(k_{a}^{12}-11D_{b}k_{d}k_{a}^{10}+45D_{b}^{2}k_{d}^{2}k_{a}^{8}-84D_{b}^{3}k_{d}^{3}k_{a}^{6}+70D_{b}^{4}k_{d}^{4}k_{a}^{4}-21D_{b}^{5}k_{d}^{5}k_{a}^{2}+D_{b}^{6}k_{d}^{6}\right)t^{7}}{2520LD_{b}^{6}}\\
 &  & -\frac{512k_{a}^{2}\left(-k_{a}^{6}+5D_{b}k_{d}k_{a}^{4}-6D_{b}^{2}k_{d}^{2}k_{a}^{2}+D_{b}^{3}k_{d}^{3}\right)\left(-k_{a}^{6}+7D_{b}k_{d}k_{a}^{4}-14D_{b}^{2}k_{d}^{2}k_{a}^{2}+7D_{b}^{3}k_{d}^{3}\right)t^{15/2}}{2027025L\sqrt{\pi}D_{b}^{13/2}}\\
 &  & -\frac{k_{a}\left(D_{b}k_{d}-k_{a}^{2}\right)\left(k_{a}^{4}-3D_{b}k_{d}k_{a}^{2}+D_{b}^{2}k_{d}^{2}\right)\left(k_{a}^{8}-9D_{b}k_{d}k_{a}^{6}+26D_{b}^{2}k_{d}^{2}k_{a}^{4}-24D_{b}^{3}k_{d}^{3}k_{a}^{2}+D_{b}^{4}k_{d}^{4}\right)t^{8}}{20160LD_{b}^{7}}\\
 &  & +\frac{1024k_{a}^{2}\left(2D_{b}k_{d}-k_{a}^{2}\right)\left(k_{a}^{4}-4D_{b}k_{d}k_{a}^{2}+2D_{b}^{2}k_{d}^{2}\right)\left(k_{a}^{8}-8D_{b}k_{d}k_{a}^{6}+20D_{b}^{2}k_{d}^{2}k_{a}^{4}-16D_{b}^{3}k_{d}^{3}k_{a}^{2}+2D_{b}^{4}k_{d}^{4}\right)t^{17/2}}{34459425L\sqrt{\pi}D_{b}^{15/2}}\\
 &  & +\frac{k_{a}\left(k_{a}^{16}-15D_{b}k_{d}k_{a}^{14}+91D_{b}^{2}k_{d}^{2}k_{a}^{12}-286D_{b}^{3}k_{d}^{3}k_{a}^{10}+495D_{b}^{4}k_{d}^{4}k_{a}^{8}-462D_{b}^{5}k_{d}^{5}k_{a}^{6}+210D_{b}^{6}k_{d}^{6}k_{a}^{4}-36D_{b}^{7}k_{d}^{7}k_{a}^{2}+D_{b}^{8}k_{d}^{8}\right)t^{9}}{181440LD_{b}^{8}}\\
 &  & -\frac{2048k_{a}^{2}\left(D_{b}k_{d}-k_{a}^{2}\right)\left(3D_{b}k_{d}-k_{a}^{2}\right)\left(-k_{a}^{6}+6D_{b}k_{d}k_{a}^{4}-9D_{b}^{2}k_{d}^{2}k_{a}^{2}+D_{b}^{3}k_{d}^{3}\right)\left(-k_{a}^{6}+6D_{b}k_{d}k_{a}^{4}-9D_{b}^{2}k_{d}^{2}k_{a}^{2}+3D_{b}^{3}k_{d}^{3}\right)t^{19/2}}{654729075L\sqrt{\pi}D_{b}^{17/2}}  \\
 & & +\mathcal{O}(t^{10}),
\end{eqnarray*}
\normalsize
where $D_b$ is the bulk diffusion coefficient, $k_a$ and $k_d$ are the adsorption and desorption rates, $L$ is the distance between the confining walls, and $f=(1+2k_a/k_dL)^{-1}$.
